# Effects of Tip–Nanotube Interactions on Atomic Force Microscopy Imaging of Carbon Nanotubes

Rouholla Alizadegan[1], Albert D. Liao[2,3,4], Feng Xiong[2,3,4], Eric Pop[2,3,4,*], and K. Jimmy Hsia[1,2,5,*]

[1]*Dept. of Mechanical Science & Engineering, University of Illinois at Urbana-Champaign, Illinois 61801, USA*

[2]*Micro- and Nanotechnology Laboratory, University of Illinois at Urbana-Champaign, Illinois 61801, USA*

[3]*Dept. of Electrical and Computer Engineering, University of Illinois at Urbana-Champaign, Illinois 61801, USA*

[4]*Beckman Institute, University of Illinois at Urbana-Champaign, Illinois 61801, USA*

[5]*Dept. of Bioengineering, University of Illinois at Urbana-Champaign, Illinois 61801, USA*

**ABSTRACT:** We examine the effect of van der Waals (vdW) interactions between atomic force microscope (AFM) tips and individual carbon nanotubes (CNTs) supported on $SiO_2$. Molecular dynamics (MD) simulations reveal how CNTs deform during AFM measurement, irrespective of the AFM tip material. The apparent height of a single- (double-) walled CNT can be used to estimate its diameter up to ~2 nm (~3 nm), but for larger diameters the CNT cross-section is no longer circular. Our simulations were compared against CNT dimensions obtained from AFM measurements and resonant Raman spectroscopy, with good agreement for the smaller CNT diameters. In general, AFM measurements of large-diameter CNTs must be interpreted with care, but the reliability of the approach is improved if knowledge of the number of CNT walls is available, or if additional verification (e.g. by optical techniques) can be obtained.

**KEYWORDS:** Carbon nanotube (CNT), atomic force microscopy (AFM), tip–nanotube interaction, diameter.

* Corresponding authors: kjhsia@illinois.edu, epop@illinois.edu

## 1. Introduction

Carbon nanotubes (CNTs) have generated much interest due to their extraordinary mechanical and electrical properties. Numerous efforts have used CNTs as building blocks of nanoelectronic devices such as transistors, switches and wires. Unlike semiconductor nanowires, CNTs do not have sharp edges or dangling bonds, thus presenting much higher carrier mobility and thermal conductivity [1−3]. CNTs can also be used as building blocks in nanosensors, nanocomposites and other nano-electro-mechanical (NEMS) devices and nanostructures [4−7]. The CNT diameter strongly affects many of its properties such as band gap [8], carrier mobility [9], contact resistance [10], and stiffness [11]. The most common method of measuring the diameter of a CNT on a substrate is by atomic force microscopy (AFM), and the most common substrate used in practice is amorphous $SiO_2$, thermally grown on Si wafers [12−17].

AFM measurements determine the CNT diameter by sampling the "height" of the CNT on the substrate; however when placed on $SiO_2$ or other solid substrates, the CNT cross-section can deform due to van der Waals (vdW) interactions between the carbon atoms and the substrate [3, 12, 18−20]. Thus, a central question in such measurements is how to interpret the AFM topography of individual CNTs [21−24], while taking into account the deformation of CNTs due to vdW interactions both with the substrate and with the AFM tip. This work focuses on answering this important issue through detailed molecular dynamics simulations and comparison with experimental measurements. The results obtained herein should be helpful in interpreting AFM measurements of CNTs and in designing and properly evaluating future CNT-based devices.

## 2. Simulation method

To study the deformation of CNTs due to various interactions we conducted molecular dynamics (MD) simulations. Several different types of interactions were considered: the covalent bonding between the carbon atoms in CNTs, the vdW interactions between the CNT and the substrate, and the vdW interactions between the CNT and the AFM tip. The neighboring carbon atoms within CNTs form covalent $sp^2$ C−C bonds, modeled with the empirical bond order Tersoff−Brenner potential [25] which has been well characterized and extensively used by many researchers [26−29]. The C−C interactions are taken to occur only when the atoms are within the cut-off radius of each other, taken as $r_c \approx 2.5$ Å in our simulations.

When CNTs are placed on a substrate, they deform due to the vdW interactions between the carbon atoms and the SiO$_2$ substrate material. As a result, their cross-sectional profile no longer remains circular, as found in several previous studies [3, 12, 18−20]. The CNT−SiO$_2$ substrate vdW potential used in our calculations is derived elsewhere [20], but repeated here for completeness:

$$V_{vdW} = \sum_{i=Si,O} \frac{2\pi\varepsilon_i\sigma_i^3 n_i}{45} \left[ 2\left(\frac{\sigma_i}{h}\right)^9 - 15\left(\frac{\sigma_i}{h}\right)^3 \right] \tag{1}$$

where [30]:

$$\varepsilon_{Si} = 8.909 \,\text{meV}, \, \sigma_{Si} = 3.326 \,\text{Å}, \, n_{Si} = 0.0227 \,\text{Å}^{-3}$$
$$\varepsilon_{O} = 3.442 \,\text{meV}, \, \sigma_{O} = 3.001 \,\text{Å}, \, n_{O} = 0.0454 \,\text{Å}^{-3} \tag{2}$$

there should be gaps between numbers and units; $^{-3}$ should be $^{-3}$

Furthermore, the non-neighboring carbon atoms in the CNTs that are outside the cut-off radius of the covalent bond may still interact with each other by vdW forces when the interatomic distance is sufficiently close. We model such interactions via a standard Lennard-Jones (LJ) potential

$$V = 4\varepsilon \left[ \left(\frac{\sigma}{r}\right)^{12} - \left(\frac{\sigma}{r}\right)^6 \right] \tag{3}$$

with the following parameters [31]:

$$\varepsilon_C = 2.39 \,\text{meV}, \, \sigma_C = 3.415 \,\text{Å}, \, n_C = 0.176 \,\text{Å}^{-3} . \tag{4}$$

The equilibrium cross-section of several single-walled CNTs (SWCNTs) on SiO$_2$ is shown in Fig. 1(a), where each ring represents the cross-section of a single-walled CNT of a particular diameter. The inner (blue) rings represent the equilibrium shapes of CNTs with diameters $D < 2.2$ nm. In this regime, we find that the elastic energy due to relatively large intrinsic curvature of the CNTs is dominant. The outer (red) rings represent the equilibrium shapes of CNTs with diameters $D > 2.2$ nm. In this regime, the CNTs deform more as seen by their flattened cross-sections. For such large diameters the maximum curvature (which occurs at the left and right sides of the cross-section) is nearly constant, and the CNT shape is governed by the vdW interactions with the substrate.

The two distinct regions are also noted in Fig. 1(b), which reveals a bilinear relationship between the maximum curvature in a deformed CNT and inverse of its diameter ($1/D$). For $D > 2.2$ nm ($1/D < 0.45$ nm$^{-1}$) the maximum curvature of a CNT on SiO$_2$ substrate changes very slowly with the inverse diameter as shown by the nearly horizontal (red) line segment. However, for CNTs with $D < 2.2$ nm the maximum curvature begins to scale proportionally with $1/D$ (slanted blue line), indicating that CNTs on SiO$_2$ maintain a nearly circular cross-section in this regime. This bilinear dependence in turn gives rise to a bilinear relationship between the height (and to a lesser degree the width) and the diameter of the CNTs, as shown in Fig. 1(c). The width is defined here as the longest possible inscribed horizontal line within a CNT. The width changes almost linearly with the CNT diameter, whereas the height is distinctively bilinear with respect to $D$. Similar behaviors are predicted by our simulation results for a double-walled CNT on SiO$_2$ substrates as shown in Fig. 1(d), with the "deflection point" occurring at a larger CNT diameter.

In order to compare the calculated CNT profiles with experimental results measured by AFM, we then calculate the CNT cross-sections as an AFM tip of given radius is moved across the CNT. Since the radius of curvature of AFM tips (typical range $R_T \approx 2$–20 nm) is usually comparable to or larger than the CNT diameter ($D \approx 0.7$–4 nm for single-walled CNTs), the trace profile of the AFM tip will not conform to the shape of CNTs. Instead, the trace profile of the AFM tip will in general be of roughly the same height ($H$) as that of the CNT but much larger in width ($W_{AFM}$). If one does not take into account the CNT–AFM tip interaction and its resulting deformation of the CNT (i.e., for perfectly rigid CNTs), such AFM tip trace profiles can be obtained analytically based on simple geometrical considerations [32], e.g. $W_{AFM} \approx 2(2R_T h)^{1/2}$.

In experiments, the AFM tip interacts with the CNT as it moves across it through vdW forces. Such interactions, in turn, cause deformation in the CNTs being imaged, in both contact and tapping modes [33]. To determine the relationship between the AFM tip trace profile and the true dimensions of a CNT, we conduct MD simulations by taking into account the full interactions of the AFM tip and the CNT. We model the tip–CNT interaction with a simplified vdW potential, i.e. the standard Lennard-Jones potential as in Eq. (2) where $\varepsilon$ is an energy parameter, $\sigma$ is a length parameter given in Eqs. (2) and (4), and $r$ is the interatomic distance. We consider parabolic-shaped diamond AFM tips, but similar conclusions can be reached for AFM tips of different shapes and materials. If the parabolic AFM tip is defined by [32] $f(x,y) = (x^2 + y^2)/2R_T$, the collective vdW interaction per C atom can be approximated as:

$$V_{vdW} = \int_{-\infty}^{\infty} \int_{-\infty}^{\infty} \int_{f(x,y)}^{\infty} 4\varepsilon n \left[ \left( \frac{\sigma}{r} \right)^{12} - \left( \frac{\sigma}{r} \right)^{6} \right] dz dx dy \qquad (5)$$

where

$$r = \sqrt{(x-k)^2 + y^2 + (z+h)^2} \qquad (6)$$

and $(k, h)$ are the horizontal distance and depth of C atoms under the AFM tip, respectively ($h$ and $k = 0$ at the apex of the AFM tip). We note that this integral tends to give lower estimates of the total interaction potential, ignoring local spikes of closely positioned atoms. This integral cannot be evaluated analytically and should be treated numerically. However, if the lower bound of the $z$ integral is changed to zero, i.e. a flat AFM tip, an analytical formula exists [20]:

$$V_{vdW} = \frac{2\pi\varepsilon\sigma^3 n}{45} \left[ 2 \left( \frac{\sigma}{h} \right)^{9} - 15 \left( \frac{\sigma}{h} \right)^{3} \right]. \qquad (7)$$

This result indicates that in general the potential is highly localized and is non-zero only in the close vicinity of the AFM surface. The integral in Eq. (5) may be evaluated numerically via partitioning the integration domain into volumetric elements and carrying out a standard Gauss integration as shown in the inset of Fig. 2. The inset of Fig. 2 plots the collective van der Waals potential map around an AFM tip on a cross section, i.e., it is a color-contour of $V_{vdW}/4\epsilon n$ where $V_{vdW}$ is obtained via Eq. (5). This indeed shows that the integral is non-zero only in a narrow band around the tip surface. This observation allows us to use an approximate, localized potential between AFM tips and CNTs for the simulations in this work, with the form:

$$V_{vdW} \approx 4\varepsilon n \left[ A \left( \frac{\sigma}{h} \right)^{\alpha} - B \left( \frac{\sigma}{h} \right)^{\beta} \right] \qquad (8)$$

where $A = 1.645 \text{ Å}^3$, $B = 14 \text{ Å}^3$, $\alpha = 9$ and $\beta = 3.288$ are fitting parameters chosen to match the numerical data, as shown by the fit in Fig. 2 for diamond AFM tips. In this model, $h$ is the closest perpendicular distance to the surface of the AFM tip from any point in space and $n$ is the atomic density of diamond. If instead of diamond we consider silicon AFM tips, the parameters are [34]:

$$\varepsilon_{Si} = 6.46 \text{ meV}, \ \sigma_{Si} = 3.621 \text{ Å}, \ n_{Si} = 0.05 \text{ Å}^{-3} \text{ there should be gaps between num-} \qquad (9)$$

bers and units; $^{-3}$ should be $^{-3}$

with the parameters of the approximation becoming $A = 1.943$ Å$^3$, $B = 16.49$ Å$^3$, $\alpha = 9$ and $\beta = 3.174$. These *effective potentials* for silicon and diamond tips are employed throughout this work, dramatically increasing the efficiency of the simulations.

In both experiment and simulations we need to apply a small downward compressive force on the tip at all times to ensure the stability of the AFM scan. When scanning a solid surface the magnitude of this force is insignificant. However, for deformable surfaces such as CNTs, the magnitude of the compressive force will affect the result. In the current simulations, we apply a compressive force ranging from 0.1 nN (extremely small) to 10 nN. It should be noted that these forces are small, such that the deformations of the substrate and tip are negligible in our simulations. Additional, exhaustive discussion of our AFM parameters is given in the Electronic Supplementary Material (ESM).

## 3. Simulation results

Figure 3 shows typical simulations of single-walled CNT shapes at different stages as the AFM tip moves across their surface. For illustrative purposes we choose a large (45,45) SWCNT (~6.17 nm in diameter) on the SiO$_2$ substrate, which is the upper diameter limit at which SWCNTs have been fabricated and observed [35, 36]. Figure 3(a) shows the shape of a SWCNT when the AFM tip is sufficiently far away such that the interactions between the AFM tip and CNT are negligible. Figures 3(b)–3(d) depict the CNT shape as the AFM tip moves across it (also see the detailed shapes of the deformed CNT and the movies in the ESM). These pictures clearly show that, because of the vdW interactions between the AFM tip and CNT as well as the deformability of the CNT, an indentation forms in the CNT near the location of the AFM tip. The degree of the indentation depends on the magnitude of the compressive force on the AFM tip and the diameter of the CNT. However, it is noted that, even with zero compressive force on the AFM tip, an indentation still forms if the CNT is sufficiently flexible, i.e. of large enough diameter. This indentation occurs because the attractive vdW interaction between carbon atoms and the AFM tip can draw the tip "into" the CNT until the elastic resistance of the tube balances the attractive force. Such attractive interactions are evident in Figs. S-1(b) and S-1(d) and in the movies (in the ESM) at the moment when the AFM tip just starts to contact the CNT and when it departs from the CNT.

Figure 4(a) shows the trace profiles of several AFM tips with different radii over a (20,20) CNT of diameter $D \approx 2.7$ nm when the tip–CNT interaction is not accounted for (i.e., a perfectly rigid CNT). Here we assume that the AFM tip is in perfect contact with the CNT as well as with the substrate, i.e., the distance between them is zero. It is clear that the height "measured" by the AFM tip trace profile, $H$, can accurately represent the true height of the CNT. However the width "measured" by the AFM tip trace profile, $W_{\mathrm{AFM}}$ is much larger than the true width of the CNT, $W$. Although highly simplified, this result nevertheless serves to illustrate that the measured width of CNTs from AFM images cannot be used to characterize the dimension of CNTs.

Figure 4(b) shows the trace profile of a diamond AFM tip with radius $R_T = 70$ Å as it moves across a (20,20) SWCNT, perpendicular to its longitudinal axis with and without the vdW interactions. These simulation results indicate that the height of a CNT measured by AFM is smaller (or slightly larger for small diameter SWCNTs, see Fig. 5) than its true height, and the measured width of a CNT is significantly larger than its true width. Using the results in Figs. 4(b) and 1(c), we can in principle determine the true diameter of the CNT from the AFM measurement data, at least for the idealized case of a perfectly flat substrate and well-defined AFM tip shape.

Figure 4(c) shows the effect of the compressive force on the magnitude of (20,20) CNT deformation along with some of the deformed intermediate configurations of these nanotubes. Since large diameter SWCNTs are extremely compliant in the radial direction, small compressive forces can cause large radial deformations and may even lead to full collapse, as discussed further below.

Using numerical simulations, we can now quantitatively examine the relationship between the CNT height measured by AFM and the true CNT diameter. Figure 5(a) compares the simulated results for AFM measurements of single-walled CNTs heights on a perfectly flat silica substrate. In these simulations, the equilibrium distance between the AFM tip and the substrate is assumed to be 2.7 Å which is the contact distance of carbon atoms on a silica substrate from our model. The actual height of a CNT on substrate is a monotonically increasing function of the nanotube diameter. However, when the CNT–AFM tip vdW interaction is taken into account, the apparent height of the CNT measured from the AFM tip trace is no longer the true CNT height. In particular, for CNTs with $D <$ ~2.0 nm, the AFM measurements should be very close to the actual CNT height; for CNTs with $D > \sim 2.0$ nm, however, the measured height data could be notably lower than the real height of the CNTs. Furthermore, there does not seem to be a one-to-one relationship between these two values, and one must be careful in interpreting the actual diameter of the CNT from AFM measurements alone, particularly for CNTs with $D > 2$ nm.

Our simulations also indicate that the AFM tip material (i.e. using a different interatomic potential) and the temperature do not play major roles in the behavior discussed above (more detailed information on this issue is provided in the ESM, see Figs. S-2 and S-4). On the other hand, the number of walls of the CNT has significant effects on the CNT height measurements. Figure 5(b) compares the true height of double-walled CNTs (DWCNTs) and the height that is measured by a diamond AFM tip. The diameter of a DWCNT is defined as the diameter of the outermost shell in the circular state. The inner shell is a CNT whose radius is $\sim 3.4$ Å (the graphite interlayer distance) smaller than the outer shell. As shown in Fig. 5(b), DWCNTs are less compliant in the transverse direction than SWCNTs, and therefore their AFM height reading is expected to be closer to the true diameter of these nanotubes for $D < \sim 3.0$ nm. Consequently, we expect the accuracy of diameter measurements from AFM "height" to be further improved for multi-walled CNTs (MWCNTs) with three or more shells, which suffer less deformation when placed on solid substrates. Our simulation results also reveal that there may be a relative rotation between the two shells of the DWCNTs as the AFM tip moves across the tube (see movies in the ESM). Indeed in a novel experiment, DeBorde et al. devised a technique to identify individual SWCNTs and DWCNTs based on AFM measurements of height vs. downward compression force [33].

Figure 5(c) compares the results of AFM tip measurements of the width of different SWCNTs over a perfectly flat silica substrate. The true width of a CNT is almost a linear function of the nanotube diameter. The apparent width of the CNT is highly dependent on the radius and shape of the AFM tip, but in general, it is much larger than the true CNT width. However, because of the one-to-one relationship between these two quantities, such measurements may be used to estimate the CNT dimensions from the AFM width measurements. To obtain truly accurate information about CNTs from AFM measurements, however, it is necessary to have the height and width measured by AFM as well as knowledge of the AFM tip radius and the number of walls of the CNT. The number of walls of CNTs may be obtained, for instance, by the experimental nano-indentation method employed in Ref. [37]. However, in practice, such complete *a* *priori* knowledge is generally unavailable, and thus the AFM measurements of CNTs must be carefully interpreted.

Finally, Fig. 5(d) illustrates the quantitative variation of the measured height of CNTs as a function of compressive force on the SWCNT. It is evident from these curves that compressive force has an important role in the height profile measurements, and it can further complicate the interpretation of AFM height measurements. The height vs. diameter relationship is highly non-linear in the presence of downward loads, and therefore, the height data cannot be effectively utilized to deduce diameter information under these conditions. According to our simulations, downward forces as small as 5 nN can cause the CNTs to locally collapse under the AFM tip if the CNT diameter is larger than a certain value. In very large diameter SWCNTs, the collapse may propagate along the entire length of the nanotube. There could also be dynamical effects of the tapping mode as well as the effect of air, photoresist and substrate roughness that also need to be taken into account. These effects are not readily accessible for theoretical quantification, but as discussed in the subsequent sections, they may play a role in AFM experiments.

## 4. Comparison with experiments

We compared our numerical results with experimental measurements of CNTs on $SiO_2$ by AFM and by resonant Raman spectroscopy, another method used to estimate CNT diameters in practice [38–41]. Single-walled CNTs were grown by chemical vapor deposition (CVD) from Fe catalyst on an $SiO_2$(90 nm)/Si substrate and electrically connected with Ti/Pd contacts [42]. The height of the CNTs on the substrate was measured using an Asylum AFM with a silicon tip of radius ~10 nm, in air. The tapping mode scan was performed with a scan speed of 2 μm/s and a scan size of 1 μm x 1 μm [34]. A typical image for one CNT is shown in Fig. 6(a). There is some error to this measurement coming from the surface roughness of the oxide substrate ($\Delta \sim 3$ Å). The CNT height values were obtained by taking averages from seven scan readings at different regions of the CNT. Both AFM and Raman measurements were carried out after the CNTs had been annealed at 300 $^oC$ in vacuum (~$10^{-5}$ torr) for >1 hour to remove surface contaminants such as residual photoresist. Nevertheless, some surface adsorbates cannot be avoided as the AFM and Raman measurements were all performed in room-temperature ambient air.

For comparison, the actual diameters of several single-walled CNTs were also measured using resonant Raman spectroscopy with a He–Ne laser wavelength of 633 nm. A typical scan setting is 16 accumulations of 90 sec long scans using ~2 mW power. Using the radial breathing mode (RBM) peak shown in Fig. 6(b), the diameter of the CNT can be obtained from

$$\omega_{RBM} = (A/D)\sqrt{1 + C_e D^2} \qquad (10)$$

where $C_e = 0.065 \ nm^{-2}$ accounts for the environmental effect of the $SiO_2$ substrate [39, 41], $A = 227 \ cm^{-1}nm$ is the proportionality constant from elasticity theory, $D$ is in nm, and $\omega_{RBM}$ is in $cm^{-1}$. The Raman G peak (discussed below) was also used as an alternative approach to estimate the CNT diameter. Figure 7 compares the CNT heights measured by AFM with the diameters estimated from the Raman measurements, along with additional values from the literature [43, 44]. In general, we note that the AFM measurements tend to under-predict the actual diameter of the CNT, as suggested by the simulations in the previous sections.

## 5. Discussion and recommendations

The results of this work suggest that, while AFM measurements remain the most convenient method for estimating the diameters of substrate-supported CNTs, the numerical data must be carefully interpreted. For single-walled CNTs supported by an atomically-smooth substrate the CNT diameter information obtained from AFM is generally expected to be reliable up to diameters $D \sim 2$ nm, whereas for double-walled CNTs (which are stiffer) this limit is extended up to $D \sim 3$ nm. An uncertainty of 0.3–0.8 nm exists around these ranges due to $SiO_2$ roughness and surface contamination as described below.

In practice, AFM measurements on CNTs are most commonly performed in air, where water, photoresist, and other surface contaminants can alter the measured results (and are difficult to include in our MD simulations). Even if these were eliminated, the effect of substrate surface roughness ($\Delta \sim 3$ Å for the amorphous $SiO_2$ substrate tested above) lends some uncertainty to the measured diameters. For instance, Ishigami et al. [45] found that the single-layer graphene height fluctuation on $SiO_2$ as measured by AFM is ~8 Å before and ~3 Å after cleaning of photoresist residue in an argon/hydrogen atmosphere at 400 °C. Moreover, the imaged height of a graphene layer on $SiO_2$ was ~9 Å in air and ~4.2 Å in vacuum, with the latter more closely comparable to the interlayer spacing in graphite of 3.4 Å.

While graphene tends to adhere [45] to the $SiO_2$ substrate and follow its corrugations, CNTs could span between microscopic valleys and hills, causing them to become "more circu-lar" (and less flattened) than what has been calculated in this paper with a perfectly smooth substrate. There are also experimental reports [46] of the effect of a water meniscus on the behavior of AFMs and nanotubes which can affect the diameter measurement and interpretation. Such effects are difficult to model in a simple, generalized approach such as that presented here, but ought to be kept in mind as additional challenges in the interpretation of experimental data. However, we believe the computational results of this work provide insights into the mechanisms at play in AFM microscopy of CNTs on solid substrates and should lead to a better understanding for the design and analysis of carbon-based nanodevices and nanostructures.

Finally, we wish to briefly comment on approaches that are alternative or complementary to AFM for obtaining the true diameter of CNT devices. Among these, transmission electron microscopy (TEM) has been used to measure the diameter and number of CNT walls beginning with the initial studies of Iijima [47]. The challenge of TEM is that CNTs must be placed on ultra-thin, electronically transparent membranes, and it cannot be directly applied to typical CNT devices on common $SiO_2$/Si substrates. Nevertheless, recent work has succeeded in combining in situ TEM and device studies [48] for some specialized test structures. Scanning tunneling microscopy (STM) [49] can also be used to obtain the diameter of individual CNTs. However STM requires samples to be placed on a conducting substrate in ultra-high vacuum, thus preventing direct application to CNT devices which are typically on $SiO_2$, sapphire or quartz. The STM measurements may also suffer from the same tip–sample interaction studied in this paper, as well as additional geometrical effects on the tunneling probabilities [50].

Optical measurements offer another alternative, non-invasive approach for obtaining CNT diameters. These include three methods, photoluminescence (PL) [51, 52], Rayleigh scattering [53, 54], and resonant Raman spectroscopy [39, 55]. The restriction of PL is that it can only be used to measure semiconducting but not metallic CNTs. Rayleigh scattering and resonant Raman spectroscopy have the added benefit that they can be used to detect both metallic and semiconducting CNTs. The drawback of Rayleigh scattering is that, for individual CNT detection, environmental perturbations from the substrate significantly lower the signal to noise ratio, which requires the samples to be suspended, an undesirable configuration for device studies. Finally, Raman spectroscopy requires the CNT to be resonant with the laser excitation energy, but does not require the removal or special preparation of the substrate. In our studies described above, only approximately 1 in 10 CNTs displayed sufficient resonance with the laser energy to reveal a radial breathing mode (RBM). Such Raman measurements of RBMs are also limited by the cut-off filter present in Raman systems, allowing only detection of $\omega_{RBM} > 100$ cm$^{-1}$ and thus restricting the measurable CNT diameter range to $< 2.78$ nm. In addition to the RBM, the G peak splitting between G$^+$ and G$^-$ modes also exhibits diameter dependence. The G$^-$ mode Raman shift is given by the empirical relationship $\omega_{G^-} = B - C/D^2$ cm$^{-1}$ where $D$ is in nm and the constant $B = 1591$ cm$^{-1}$ refers to the unshifted G$^+$ mode. Here $C$ is determined by the chirality (semiconducting or metallic) of the CNT. In our estimates we used the following values from the literature: $C = 47.7$ cm$^{-1}$nm$^2$ for semiconducting CNTs, and $C = 79.5$ cm$^{-1}$nm$^2$ for metallic CNTs [40]. However, while this relationship works well for semiconducting CNTs, it does not work as well for metallic CNTs, because of increased sensitivity to doping [56], as can be seen in Table 1.

## Acknowledgements

This work was supported by the NSF grants CMMI 0952565 and CMMI 1029221 (R.A. and K.J.H.) and NSF award CAREER ECCS 0954423 (E.P.), the NRI Coufal Fellowship (A.L.) and the Marco MSD Center (F.X.). The authors thank T. DeBorde and E.D. Minot for the helpful discussion on the technique to estimate the compressive force on CNTs.

## Electronic Supplementary Material:

Supplementary material providing additional details of the simulation and experimental methods as well as video clips is available in the online version of this article at http://dx.doi.org/10.1007/s12274-012-0203-8.

**Figures and Tables.**

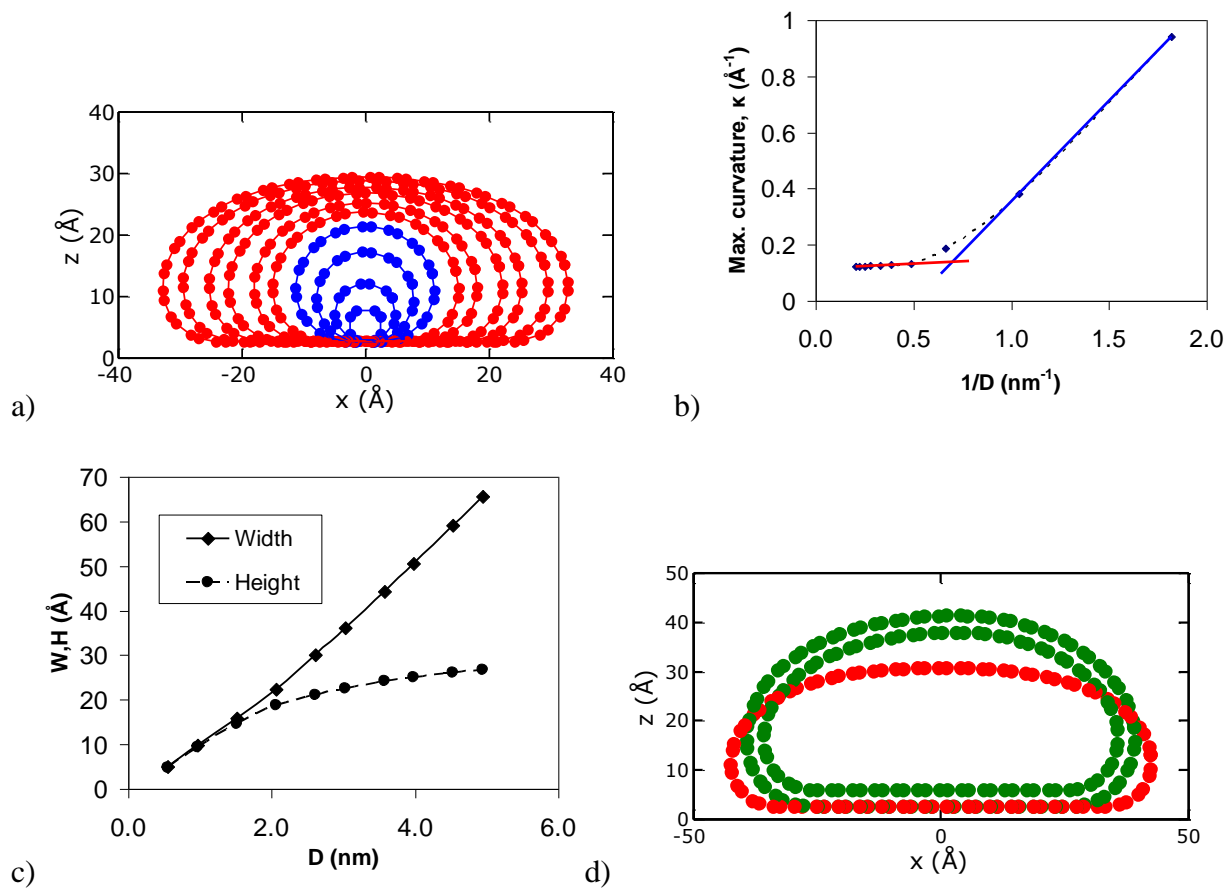

**Figure 1** CNTs on SiO$_2$ substrate: (a) Some of the typical equilibrium state single-walled CNT geometries. The inner (blue) CNTs represent the nanotubes with diameters less than 2.2 nm. The outer (red) CNTs represent the nanotubes with diameters larger than 2.2 nm. (b) Maximum curvature as a function of inverse of the tube diameter. Maximum curvature always happens at the side edges of the tubes. (c) Variation of height and width of SWCNTs on SiO$_2$ with the diameter of the nanotubes. (d) A typical (45,45) armchair SWCNT at equilibrium state (red) compared to a double-walled CNT with the same outer shell CNT (green).

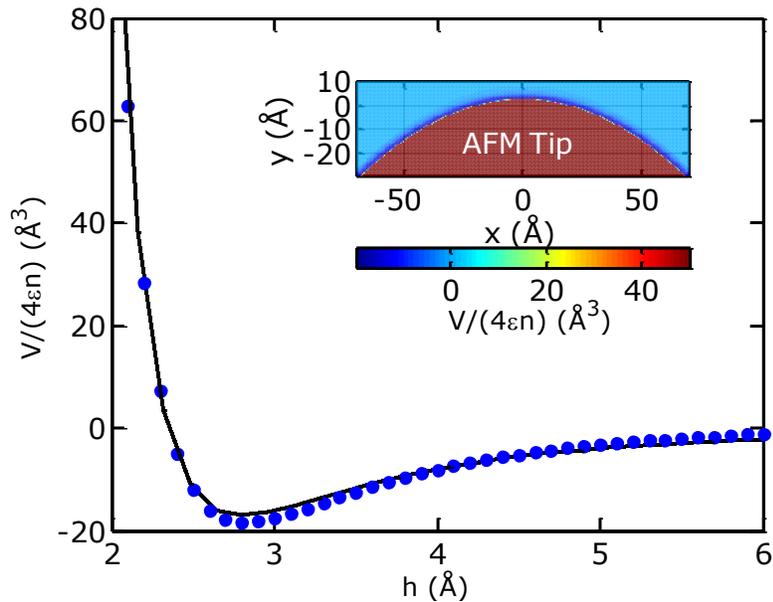

**Figure 2** Comparison of the numerical van der Waals potential (Eq. 5, the blue dots) and the analytical fit to it (Eq. 8, the solid line). (Inset) The color-contour map of $V_{vdW}/4\epsilon n$ where $V_{vdW}$ is obtained via Eq. (5) plotted around the cross section of an AFM tip. The light blue region is identically zero (after the incorporation of a cut-off radius, $r_c = 2.5\sigma_c$) and the dark magenta region is the AFM tip.

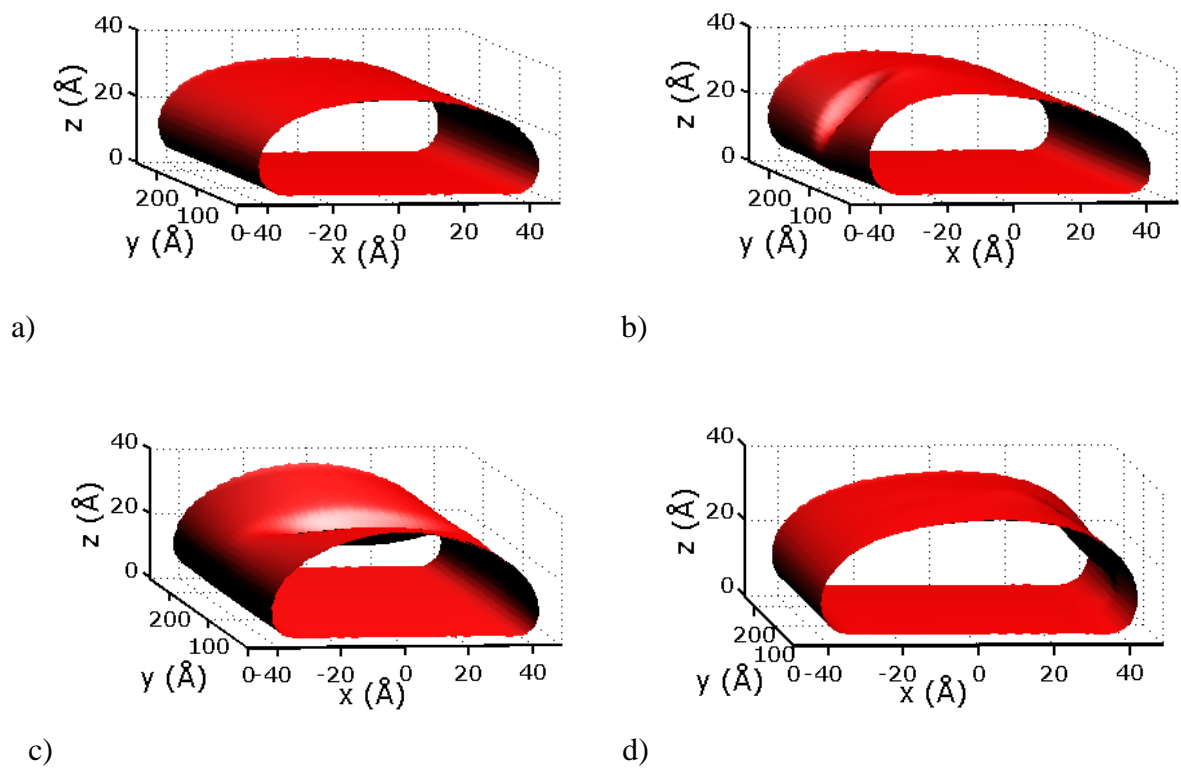

a)

b)

c)

d)

**Figure 3** The different stages of the AFM scan, here for a very large diameter CNT for better visualization (another example is shown in the ESM). A relaxed armchair (45,45) CNT (a) snaps towards the tip (b), dips in (c), and snaps away from the tip (d) during the motion of the AFM tip over it from left to right.

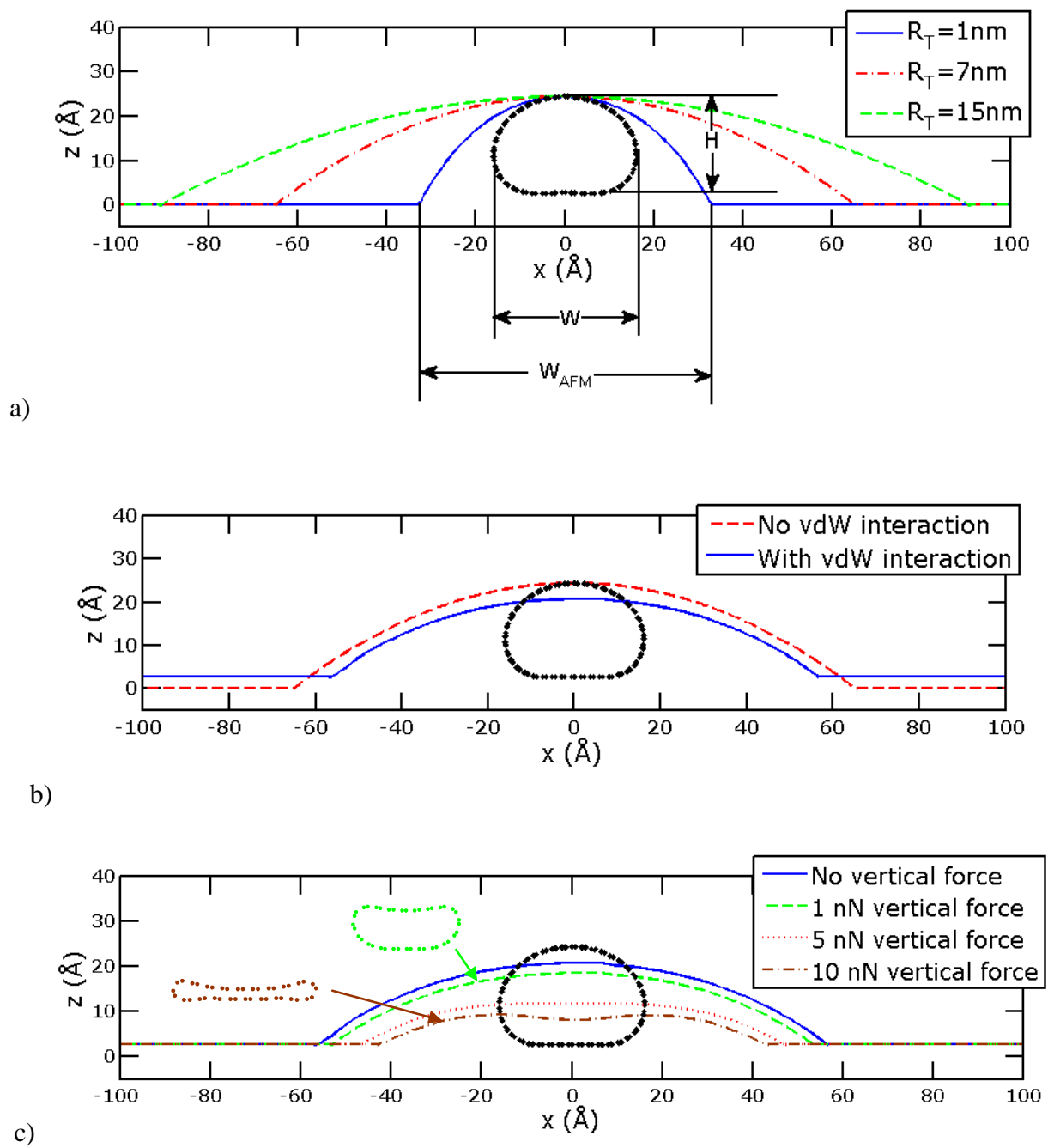

a)

b)

c)

**Figure 4** (a) The non-interacting parabolic AFM tip path for some different tip radii ($R_T$). The CNT is a typical (20,20) armchair nanotube with the actual diameter of ~2.7 nm. (b) The AFM tip path for the same nanotube with and without considering the vdW interaction. The tip radius is $R_T = 70$ Å for both curves and the black dots show the relaxed position of the CNT. (c) This figure compares the trajectory of a diamond AFM probe with a tip radius of $R_T = 70$ Å over a (20,20) CNT with the various values of downward vertical force.

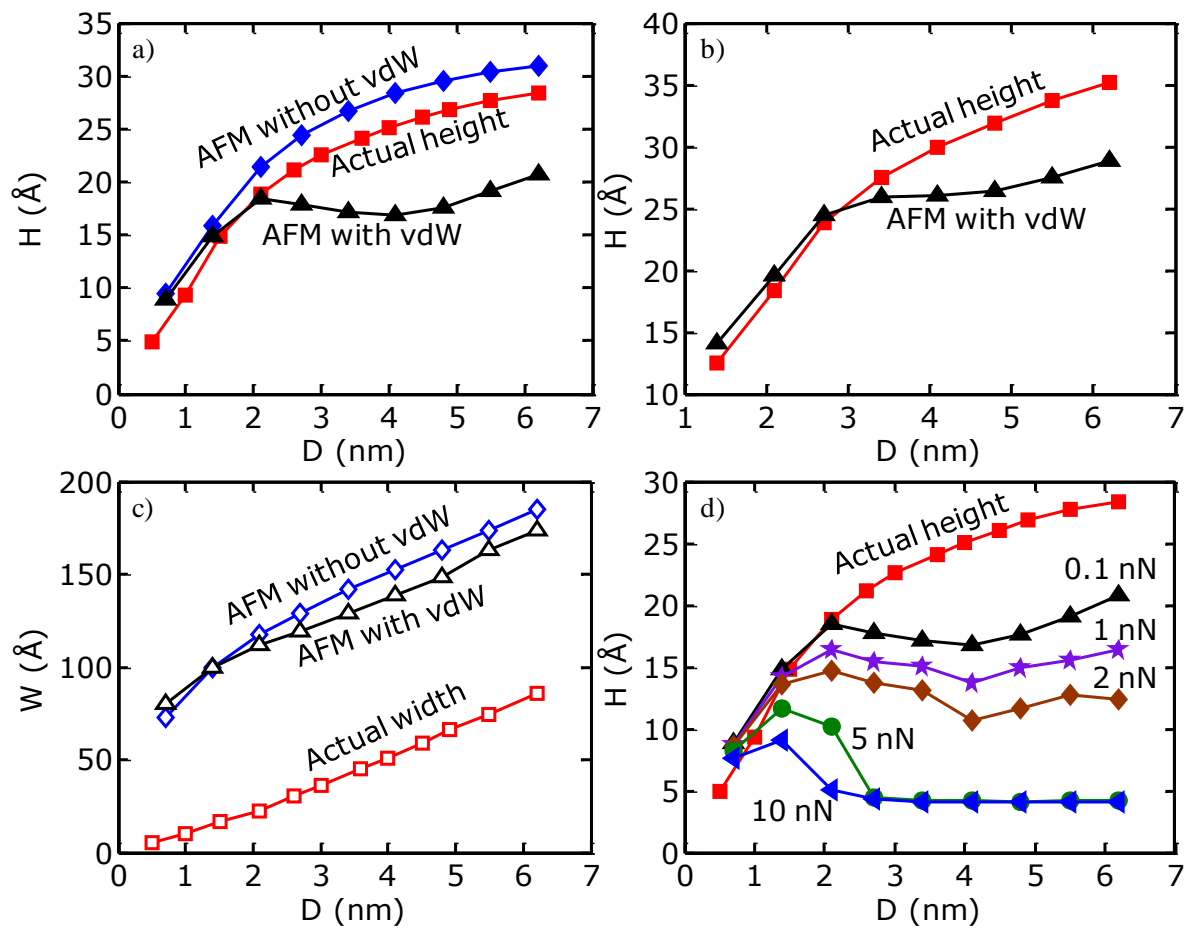

**Figure 5** (a) The variation of the height of SWCNTs on atomically flat SiO$_2$ with the nanotube diameter. The three lines are the actual height of CNTs without interactions with an AFM tip (red squares), the simulation results in the presence of an AFM tip without vdW interactions (blue diamonds) and the simulation results in the presence of an AFM tip taking into account vdW interactions (black triangles). (b) The variation of the height of DWCNTs with the nanotube diameter. (c) The variation of the width (open symbols) of SWCNTs as a function of actual nanotube diameter. (d) The effect of downward compressive force from AFM tip on the SWCNT. The tip radius is $R_T = 70$ Å for all curves. Lines are just drawn to guide the eyes.

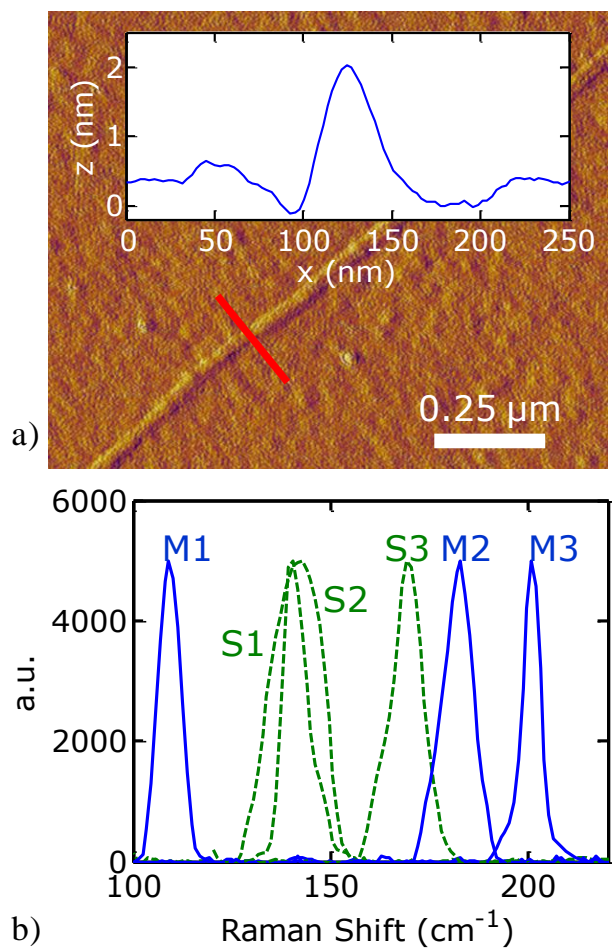

**Figure 6** (a) An AFM image of a CNT on SiO$_2$. The measured height of the CNT is 1.6 ± 0.2 nm. The height is obtained by averaging measurements at seven different locations along of the CNT, with one typical location shown by the red line and the measurement shown the inset. (b) The radial breathing mode (RBM) peaks from a resonant Raman spectroscopy measurement for six different CNTs. The solid blue lines represent metallic CNTs and the green dashed lines represent semiconducting CNTs. Curves are scaled to be the same height after having the baseline removed and smoothed.

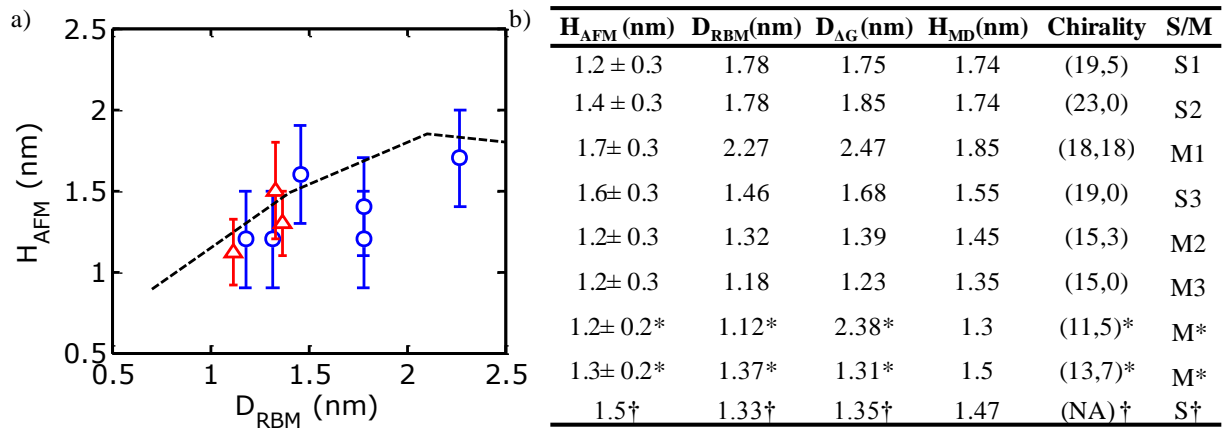

a)

b)

| $H_{AFM}$ (nm) | $D_{RBM}$ (nm) | $D_{AG}$ (nm) | $H_{MD}$ (nm) | Chirality | S/M |
|---|---|---|---|---|---|
| $1.2 \pm 0.3$ | 1.78 | 1.75 | 1.74 | (19,5) | S1 |
| $1.4 \pm 0.3$ | 1.78 | 1.85 | 1.74 | (23,0) | S2 |
| $1.7 \pm 0.3$ | 2.27 | 2.47 | 1.85 | (18,18) | M1 |
| $1.6 \pm 0.3$ | 1.46 | 1.68 | 1.55 | (19,0) | S3 |
| $1.2 \pm 0.3$ | 1.32 | 1.39 | 1.45 | (15,3) | M2 |
| $1.2 \pm 0.3$ | 1.18 | 1.23 | 1.35 | (15,0) | M3 |
| $1.2 \pm 0.2$* | 1.12* | 2.38* | 1.3 | (11,5)* | M* |
| $1.3 \pm 0.2$* | 1.37* | 1.31* | 1.5 | (13,7)* | M* |
| 1.5† | 1.33† | 1.35† | 1.47 | (NA)† | S† |

**Figure 7** (a) Single-walled CNT dimensions measured by AFM (i.e. apparent "height" $H_{AFM}$) versus those extracted from their Raman RBM measurements (see text). Blue circles indicate our data, red triangles indicate points take from the literature, and the dashed line is generated from our MD simulations (the same as in Fig. 5a with vdW interactions). (b) Table comparing the apparent height of CNTs measured by AFM ($H_{AFM}$) and simulated with molecular dynamics ($H_{MD}$) for diameters that were characterized with Raman spectroscopy using the RBM ($D_{RBM}$) and G peak splitting ($D_G$) method (see text). The estimated CNT chirality is also provided by comparing the RBM and G peak splitting diameter to the Kataura plots [41, 43, 52]. Values which are measured or generated for this study correspond to the RBM peaks from Fig. 6(b). The values from Ref. [43] and [44] are marked by * and † respectively. For CNTs with a large diameter, the AFM measurement tends to give a lower value than the more accurate diameter measured by Raman spectroscopy.